\documentclass{ws-book975x65edit} 
\usepackage{ws-book-har}

\newcommand{\beq}{\begin{equation}}
\newcommand{\eeq}{\end{equation}}
\newcommand{\beqa}{\begin{eqnarray}}
\newcommand{\eeqa}{\end{eqnarray}}
\newcommand{\om}{\Omega_m}

\begin{document} 

\bibliographystyle{ws-book-har} 

\chapter{Constraining Models of Dark Energy} 
%\title{Constraining Models of Dark Energy} 
\author{Eric V.\ Linder \\
Berkeley Lab \& University of California, Berkeley, CA 94720, USA \\ 
Institute for the Early Universe, Ewha Womans University, Seoul, Korea} 
\date{\today}

\section{Measuring Dark Energy} \label{sec:intro}

Acceleration of the cosmic expansion opens new frontiers in our 
understanding of the quantum vacuum, gravitation, high energy physics, 
extra dimensions, or some combination of these research areas.  Although 
we give the name ``dark energy'' to the accelerating mechanism, we know 
very little about its characteristics or even the area of physics from 
which it arises.  One of the great endeavors of the past decade has been 
the evaluation of different observational techniques for measuring dark 
energy properties and of theoretical techniques for constraining models
of cosmic acceleration given cosmological data.  
This chapter reviews a few of the key developments, promises, and 
cautions for revealing dark energy.  We also present a few new 
calculations, on direct detection of acceleration through redshift drift, 
the minimum uncertainty in the equation of state, and testing gravity.
For other recent reviews, see \citet{calkam,friehuttur,leon,linrop,silvestri}. 

One can probe dark energy through 1) its effect on the cosmic expansion, 
2) its indirect effect on the growth of observed structure from 
the dark energy influence on the expansion, and 3) any direct 
contribution of it to the growth of structure.  The first includes 
geometric probes and involves distances and volumes, coming directly 
from the metric.  The second and third are growth probes, involving the 
growth factor and growth rate of matter density perturbations or the 
overall gravitational potential.  The third, direct contribution of dark 
energy to growth predominantly occurs when the dark energy represents a 
modification of the gravitational equations for growth, as dark energy 
tends to be smooth on subhorizon scales in most other models and so dark 
energy density perturbations contribute little to the Poisson equation. 

The existence of dark energy was first discovered through the geometric 
probe of the distance-redshift relation of Type Ia supernovae 
\citep{perl99,riess98}.  Such data have been greatly expanded and refined 
so that now the analysis of the Union2 compilation of supernova data 
\cite{union2}, together with other probes, establishes that 
the energy density contribution of dark energy to the total energy 
density is $\Omega_{de}=0.281\pm0.017$ and the dark energy equation of 
state, or pressure to density ratio, is $w=-1.03\pm0.09$.  

Note that measurement of distances relative to low redshift, e.g.\ 
the Hubble diagram of supernovae, is the most sensitive probe of cosmic 
acceleration to date, i.e.\ it probes the deceleration parameter or 
$\ddot a$ most directly.  Recall that the luminosity distance follows 
immediately from the metric, with $d_l=a^{-1}\int dt/a=a^{-1}\int da/(a^2H)$ 
where $a$ is the expansion factor and $H=\dot a/a$ is the Hubble expansion 
rate. 

Is there a more direct measure of acceleration than the distance 
(all distances being similar to the luminosity distance), which involves 
at best $\dot a$, rather than $\ddot a$ itself?  One can devise a 
quantity involving the finite difference between expansion rates: 
the redshift drift $dz/dt_0$ as the universe ages, i.e.\ as we observe a 
given source at different epochs \citep{sandage,mcvittie,fpoc}.  The drift 
is of order the Hubble time, $\dot z\sim H$, so this is beyond present 
observational capabilities.  Since it is an intriguing idea, though, let 
us pursue it a little further.  For one thing, other effects intrude due 
to not living in a perfectly smooth universe. 

The redshift of a source seen by an observer is 
\beq 
1+z=\frac{(g_{\mu\nu}k^\mu u^\nu)_e}{(g_{\mu\nu}k^\mu u^\nu)_o}\,, 
\eeq 
where subscript $e$ denotes the emitter frame, $o$ denotes the observer 
frame, and $g_{\mu\nu}$ is the metric, $k^\mu$ is the photon 
four-momentum, and $u^\nu$ is the source or observer four-velocity. 
If we want to find $dz/dt_o$ then we must take into account three 
contributions: 1) peculiar accelerations in the form of $\dot u$, 
2) the homogeneous and inhomogeneous evolution of the metric, 
involving the scale factor $a(t)$ and the metric potentials $\psi$ 
and $\phi$, and 3) the geodesic equation of the photon through the 
gravitational potentials along the path, i.e.\ $k^\mu(x^\mu)$. 

To estimate the impact of peculiar accelerations $\dot u$, consider 
that they involve a spatial derivative of the potential, i.e.\ 
$\vec\nabla\psi\sim\dot u$.  Thus one can write the order of magnitude 
as $\dot u\sim (k/H)H\psi$, where $k$ is the characteristic inverse 
length scale.  This contribution could be of order $H$ for sources 
living in a galactic potential, just like the expansion factor 
contribution.  One might be able to reduce this ``noise'' by using an 
array of redshift drift sources across the sky. 

The metric and geodesic effects on the redshift drift are given by 
solving the geodesic equation, yielding 
\beqa 
\frac{dz}{dt_o}&=&\frac{\dot a_o-\dot a_e}{a_e}+2[\dot\psi_e-(1+z)\dot\psi_o] 
+\frac{2}{a_e}\partial_1(\psi_e-\psi_o)-(\dot\psi_e-\dot\phi_e)\nonumber\\ 
&\quad&+(1+z)(\dot\psi_o-\dot\phi_o) 
-H(z)\,(\phi_o-\phi_e)+H_0\,(1+z)[a_o k^0_o]^{(1)}\, 
\eeqa  
where $k^0_o{}^{(1)}$ is the first order correction to the observed photon 
frequency (and can be defined to be zero).  Note that some terms only 
arise in modified gravity where the metric potentials $\psi\ne\phi$.  
The first term, discussed 
below, is of order $H$; all remaining terms are at most of order $k\phi$, 
or $(k/H)\phi$ relatively, and so should be smaller by at least two orders 
of magnitude.  A possible exception might involve supermassive binary black 
holes, where $\dot\phi$ arises from inspirals; see Appendix B of 
\citet{yunes}. 

The first term is the standard McVittie-Sandage result 
\citep{mcvittie,sandage}, and involves the difference of the expansion rate 
between two redshifts, i.e.\ it provides some measure of the acceleration 
of the expansion.  Even if this could be cleanly separated and measured, 
it provides only an order unity precision measurement of the acceleration 
for observations accurate to $\delta z\sim 10^{-9}/$(10 years).  To determine 
an assumed constant equation of state $w$ to 1\% would require $\delta z\sim 
10^{-12}$/year. 
Thus direct measurement of dark energy seems infeasible in the next 
generation.  Instead we return to the geometric and growth probes.

%%%%%%%%%%%%%%%%%%%%%%%%%%%%%%%%%%%%%%%%%%%%%%%%%%%%%%%%%%%%%
\section{Current Data} \label{sec:data} 

The response of the experimental cosmology community to the dark energy 
puzzle has been gratifyingly strong and diverse.  As of mid 
2010, experiments are underway using Type Ia and Type II supernovae, 
baryon acoustic oscillations, cosmic microwave background measurements, 
weak gravitational lensing, and galaxy clusters with the Sunyaev-Zel'dovich 
effect and X-rays.  Even within the next year, more data will be released 
with continuing impact. 

This activity is valuable, especially because currently the only 
technique that by itself unambiguously detects cosmic acceleration -- 
let alone characterizes it precisely -- is Type Ia supernovae.  Other 
techniques need to be combined with external information, such as the 
Hubble constant, large scale structure data, or each other.  X-ray clusters 
\citep{vikhlinin} and weak gravitational lensing \citep{schrabback} 
see acceleration at the 1-1.5$\sigma$ level, if systematics do not enter 
in excess of the reports.  Of course when data for various probes are 
combined together, one has strong concordant evidence for acceleration 
and moderate constraints on the dark energy equation of state viewed 
in a time-averaged, i.e.\ assumed constant, sense. 

\begin{figure}[!htb]
\begin{center}
\psfig{file=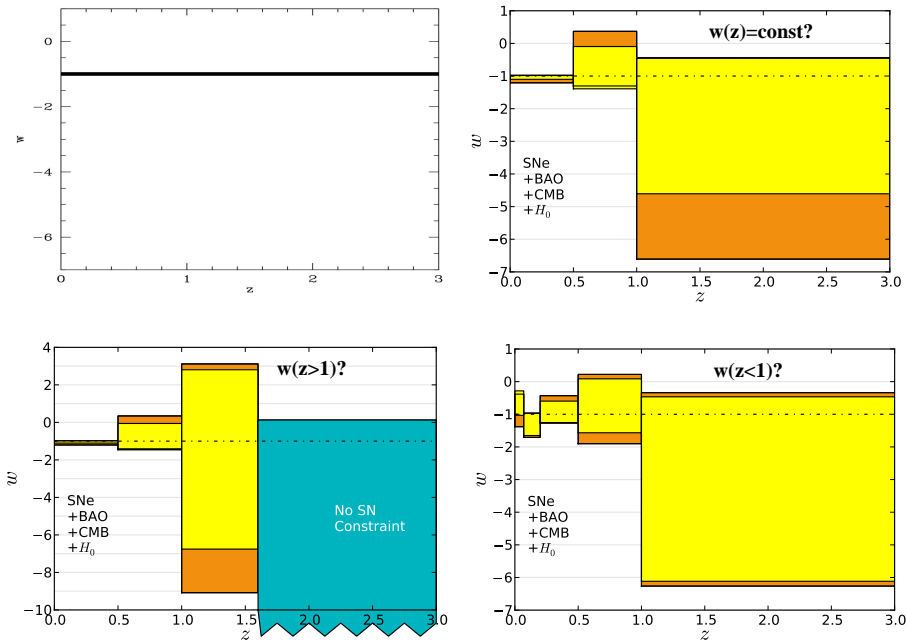,width=0.47\textwidth}
\psfig{file=wbins_3_BAO_CMB_H0wmap7.epsi,width=0.45\textwidth} \\ $\,$ \\ 
\psfig{file=wbins_3+1_BAO_CMB_H0wmap7.epsi,width=0.48\textwidth}
\psfig{file=wbins_eqerr5_BAO_CMB_H0wmap7.epsi,width=0.45\textwidth}
\caption{Constraints from the Union2 supernova compilation, 
WMAP7 CMB, SDSS DR7 baryon acoustic oscillation, and Hubble 
constant data on the dark energy equation of state $w(z)$, in 
redshift bins.  Top left plot appears to show that data 
have zeroed in on the cosmological constant value of $w=-1$, 
but this assumes $w$ is constant.  When one allows for the values 
of $w$ to be different in different redshift bins, our current 
knowledge of dark energy is seen to be far from sufficient. 
Top right plot shows that we do not yet have good constraints on 
whether $w(z)$ is constant.  Bottom left plot (note change of scale) 
shows we have little 
knowledge of dark energy behavior, or even existence, at $z>1$. 
Bottom right plot shows we have little detailed knowledge of dark energy 
behavior at $z<1$.  Outer (inner) boxes show 68\% confidence limits 
with (without) systematics.  
The results are consistent with $w=-1$, but also allow 
considerable variation in $w(z)$.  Adapted from \citet{union2}. 
}
\label{fig:union2}
\end{center}
\end{figure}

It is worthwhile to get an overview of the current results going beyond 
a constant equation of state.  All the published Type Ia supernova data 
has been brought together and uniformly analyzed, employing systematics 
tests and blinded cosmology fitting, in the Union2 compilation 
\citep{union2}.  
In addition to results in terms of a constant equation of state $w$, time 
variations $w(a)=w_0+w_a(1-a)$, $w_i(z)$ binned in redshift, and dark energy 
density $\rho_i(z)$ binned in redshift have been employed as well.  This 
has also included other cosmological probes.  All cases are consistent 
with the cosmological constant value of $w=-1$, however the results agree 
as well as with many other, dynamical models.  
Figure~\ref{fig:union2} illustrates the current state of our knowledge, 
viewed in terms of binned $w$.

Many different physical origins for acceleration are viable according to 
current data.  Several have been explored in detail for the earlier Union1 
compilation \citep{union1}, showing that physical models quite distinct 
from  the cosmological constant $\Lambda$ are acceptable \citep{beylam}.  
One of the interesting results is that for some dark energy origins not 
all probes exhibit the complementarity familiar from the 
$\Omega_m$-$\Omega_\Lambda$ or $\Omega_m$-$w$ planes. 
For example, BAO and CMB basically give degenerate information for 
the linear potential (cosmic doomsday) model.  Supernovae, because 
of their link to expansion through the metric, always give the most 
direct constraint on acceleration.

%%%%%%%%%%%%%%%%%%%%%%%%%%%%%%%%%%%%%%%%%%%%%%%%%%%%%%%%%%%%%%
\section{Future: Constraining What?} \label{sec:fut}

As Figure~\ref{fig:union2} demonstrates, we clearly need to learn 
considerably more about the nature of dark energy.  What is its dynamics, 
does it have further degrees of freedom beyond $w$ (clustering, interaction, 
etc.), how accurate is an effective description 
in terms of a few parameters for interpreting next generation observations? 

For the future, what type of constraints can we expect?  How many 
handles will we have on the physics behind cosmic acceleration?  Recall 
that originally the early cosmic acceleration of inflation was thought 
to be untestable but now we have a rich variety of observational 
signatures such as scalar perturbation tilt and running, spatial curvature, 
tensors, nonGaussianity, and topological defects.  Dark energy, 
although occurring at an epoch more amenable to direct observation, has not 
yet revealed so many tests.  

In most cases we do not expect a perturbation generation mechanism from 
the underlying physics because dark energy neither completely dominates 
the expansion nor has the acceleration ended, e.g.\ in a reheating epoch.  
On the plus side, one observational window is the expansion itself 
(inaccessible for inflation).  That is why the expansion equivalents 
of tilt and running -- $w$ and $w'$ -- play such a large role.  As we will 
see later in this section, another potential signature involves perturbations 
within the dark energy itself. 

\subsection{Limits on Measuring $w$ \label{sec:wlim}}

Many models have a region or limit in parameter space in which their 
equation of state closely approaches the cosmological constant value of 
$w=-1$.  As $w$ nears $-1$, it becomes increasingly difficult to achieve 
accurate enough observations to distinguish the equation of state from 
$-1$, and potentially recognize these models as distinct from $\Lambda$. 
Moreover, there is a theoretical difficulty as well due to the covariance 
between $w$ and other parameters entering the expansion history. 

Consider the simple flat model with constant $w$; the Hubble expansion is 
given by 
\beq 
H^2/H_0^2=\om (1+z)^3+(1-\om)\,(1+z)^{3(1+w)}\,. 
\eeq 
The error on $w$ from a measurement can be written as 
\beq 
\delta w=\frac{y^{-3(1+w)}}{3(1-\om)\,\ln y}\,
\left[\delta(H^2/H_0^2)-\delta\om\,(y^3-y^{3(1+w)})\right]\,, 
\eeq 
where $y=1+z$.  Take the idealized situation of a perfect measurement 
of $H^2/H_0^2$.  Then 
\beq 
\sigma(w)\approx (3{\rm -}5)\,\sigma(\om) 
\approx 0.016\,\frac{\sigma(\om)}{0.004}\,, 
\eeq 
for the idealized measurement at $z=0.5-1$. 

One can carry out a similar estimation regarding determination of whether 
the dark energy density $\rho_{de}$ is constant\footnote{I 
thank Bob Scherrer for suggesting that when $w\approx-1$, one might 
do this analysis in terms of $\rho_{de}$.}.  Here one finds 
\beq 
\sigma(\rho/\rho_0)\approx \frac{y^3-1}{1-\om}\,\sigma(\om) \approx 
(3{\rm -}10)\,\sigma(\om)\approx (0.01{\rm -}0.04)\,
\frac{\sigma(\om)}{0.004}\,. 
\eeq 
These expressions give practical bounds on the accuracy of distinction 
from $\Lambda$, since $\om$ (and other covariant parameters) will be 
imperfectly known.  (Not allowing for uncertainty in, e.g., $\om$ 
can give larger, bias effects.  See, e.g., Figure 6 of \citet{hlozek}.)

\subsection{Mapping Dynamics \label{sec:mapdyn}} 

Observables such as the distance-redshift relation and Hubble 
parameter-redshift relation can be used to test specific models of 
dark energy, but it is frequently useful to have a more model independent 
method of constraining dark energy properties.  A calibration relation 
exists between the dark energy equation of state value and its time 
variation that defines homogeneous families of dark energy physics 
\citep{depl0808}.  This calibration provides a physical basis for 
the $w_0$-$w_a$ parametrization devised to fit the exact solutions for 
scalar field dynamics \citep{linprl}. 

The resulting parameters $w_0$, $w_a$ give a highly accurate match to 
the observable relations of distance $d(z)$ and Hubble parameter $H(z)$, 
and the constraints imposed on them by data allow robust exploration 
of the nature of dark energy in a model independent manner. 
(Note that $w_0$, $w_a$ are thus defined in terms of this calibration 
as opposed to a fit to the unobservable $w(z)$.)  

The accuracy with 
respect to the observables for a diverse group of models is exhibited 
in Table~\ref{tab:calib}, for models near the extreme of current viability; 
less extreme models will be fit even better.  
Such 0.1\% or better accuracy is more than 
sufficient for next generation experiments.  While constant $w$ 
overcompresses (loses important information from) the expansion history 
information, $w_0$-$w_a$ faithfully preserves the information to 
better than the precision level of the data.  Attempts to use further 
parameters generically lack additional leverage on the data (e.g.\ see 
the next subsection), giving no real benefit (save in the possible 
exception of early dark energy models).  Thus, $w_0$, $w_a$ 
provide an excellent parameter space for the cosmic expansion history. 

\begin{table}[!htbp]
\begin{center}
\tbl{
Accuracy of $w_0$-$w_a$ in fitting the exact distances and
Hubble parameters for various dark energy models.  These numbers
represent global fits over all redshifts (except for the last three cases,
where the fit covers $z=0$-3, due to early dark energy).
Better fits can be found over finite redshift
ranges. From \citet{depl0808}. 
}
{
%\begin{tabular}{ l l r@{.}l r@{.}l  }
%\begin{tabular}{ l c c }
\begin{tabular*}{0.9\columnwidth} 
{@{\extracolsep{\fill}} l c c }
\\ 
\hline
Model & $\delta d/d$ & $\delta H/H$ \\ 
\hline
PNGB ($w_0=-0.85$) & 0.05\% & 0.1\% \\ 
PNGB ($w_0=-0.75$) & 0.1\% & 0.2\% \\ 
Linear Pot.\ ($w_0=-0.85$) & 0.05\% & 0.1\% \\ 
Linear Pot.\ ($w_0=-0.75$) & 0.1\% & 0.3\% \\ 
$\phi^4$ ($w_0=-0.85$) & 0.01\% & 0.04\% \\ 
$\phi^4$ ($w_0=-0.75$) & 0.02\% & 0.06\% \\ 
Braneworld ($w_0=-0.78$) & 0.03\% & 0.07\% \\ 
SUGRA ($n=2$) & 0.1\% & 0.3\% \\ 
\hline 
SUGRA ($n=11$) & 0.1\% & 0.3\% \\ 
Albrecht-Skordis ($\Omega_e=0.03$) & 0.01\% & 0.02\% \\ 
Albrecht-Skordis ($\Omega_e=0.26$) & 0.1\% & 0.4\% \\ 
\hline 
\end{tabular*}
}
%\caption{Accuracy of $w_0$-$w_a$ in fitting the exact distances and 
%Hubble parameters for various dark energy models.  These numbers 
%represent global fits over all redshifts (except for the last three cases, 
%where the fit covers $z=0$-3, due to early dark energy).  
%Better fits can be found over finite redshift 
%ranges. From \citep{depl0808}.} 
\label{tab:calib}
\end{center}
\end{table}

\subsection{Further Dynamics? \label{sec:furdyn}} 

The previous subsection showed that the two parameters of $w_0$, $w_a$ 
for the dark energy equation of state provide information more than 
sufficient to match the data of next generation experiments.  Their 
calibration of the expansion history data is accurate down to the 
$\sim0.1\%$ level.  Nevertheless, there is the temptation to look 
for information about $w(a)$ in some other form. 

Principal component analysis has been suggested as an alternate view 
of the equation of state, with each specific experiment determining the 
combinations of redshift-dependent functions carrying the most information 
(see, e.g., \citet{hutstar,depl,mhh} for discussions of the method).  
Here too, however, two parameters describe the 
vast majority of information.  If we want to distinguish a model from 
the cosmological constant, say, the $\chi^2$, or signal to noise, is 
\beq 
S/N=\left[\sum \frac{\alpha_i^2}{\sigma_i^2}\right]^{1/2}, \label{eq:snr}
\eeq 
where $\sigma_i$ is the uncertainty on the coefficient $\alpha_i$ of the 
$i$th principal component, where $w(a)-w_{\rm b}(a)=\sum_i \alpha_i\,e_i(a)$. 
(Note $\sigma_i$ itself captures no physics; it is the ratio 
$\sigma_i/\alpha_i$ that is important.) 

Each principal component contributes a certain amount to the statistical 
significance, and we can quantify how much the modes beyond the first two 
matter.  Taking a representative of the freezing class of dark energy, we 
can scan over every possible model parameter value, determine the 
corresponding $\alpha_i$ and $\sigma_i$, and weigh the contribution 
of higher modes.  We then repeat this for the thawing class, and an 
oscillating case (see \citet{depl0812} for details).  
Table~\ref{tab:sn2} lists the fraction of the total 
$S/N$ covered by the first two modes -- for the case for each dark energy 
class where higher modes contribute the {\it most\/}.  

\begin{table}[!htbp]
\begin{center}
\tbl{
Fraction of total signal-to-noise contributed by the first
two, or three, principal components for the case in each dark energy
class {\it most\/} favoring PCA high modes.  From \citet{depl0812}.
}
{
%\begin{tabular}{ l l r@{.}l r@{.}l  }
%\begin{tabular}{ l c c }
\begin{tabular*}{0.95\columnwidth} 
{@{\extracolsep{\fill}} l c c }
\\ 
\hline
Model & $(S/N)_2/(S/N)_{\rm all}$ & \ $(S/N)_3/(S/N)_{\rm all}$ \\ 
\hline
Freezing ($w_a=0.7$)& 0.972& \ 0.995 \\ 
Thawing ($w_a=-0.5$) & 0.997& \ 0.9998 \\ 
Oscillating ($A=0.5$) & 0.862& \ 0.922 \\ 
\hline 
\end{tabular*}
}
%\caption{Fraction of total signal-to-noise contributed by the first 
%two, or three, principal components for the case in each dark energy 
%class {\it most\/} favoring PCA high modes \citep{depl0812}.} 
\label{tab:sn2}
\end{center}
\end{table}

For the thawing 
class the higher modes contribute less than 0.3\% to the total, and for 
the freezing class less than 2.8\% in the most sensitive case, dropping 
to less than 0.5\% for modes above the third.  If we allow the oscillating 
case to reach $w=0$, then the additional contribution can be up to 14\% 
(actually much less because appropriate marginalization over the equation 
of state beyond where the distance data lie reduces this by a factor of 
several).  These are the {\it upper\/} limits on contribution by modes 
beyond the first two, for a highly idealized experiment with 0.3\% 
distance determination over $z=0-3$.  This is not to say that modes 
beyond the first two cannot contribute $S/N>1$, only that the vast 
majority of the information contained in the data comes from two modes. 
This information is found to be at essentially the same level as from 
using $w_0$, $w_a$ directly.

\subsection{Microphysics \label{sec:sound}} 

Although the equation of state information is represented accurately 
by $w_0$, $w_a$, there is further information on the dark energy 
nature that can be extracted.  Even a perfectly measured $w(z)$ does 
not generically tell us whether dark energy arises from a canonical, minimally 
coupled scalar field, a more complicated fluid description, or modification 
of gravitational theory on large scales.  The properties of the perturbations 
to the dark energy, which must exist unless it is simply a cosmological 
constant, do carry such extra information.  We can consider this 
information in terms of the sound speed. 

A dark energy sound speed below the speed of light enhances the spatial 
variations of the dark energy.  Clustering dark energy influences the 
growth of density fluctuations in the matter, and the pattern of large 
scale structure, and an evolving gravitational potential generates the 
Integrated Sachs-Wolfe (ISW) effect in the cosmic microwave background. 
These effects are suppressed while the equation of state is near $-1$, 
so knowledge of the sound speed is strongest for models that have 
a period where $w$ is far from $-1$, and in particular for early dark energy 
models. 

\begin{figure}[!htb]
  \begin{center} 
\psfig{file=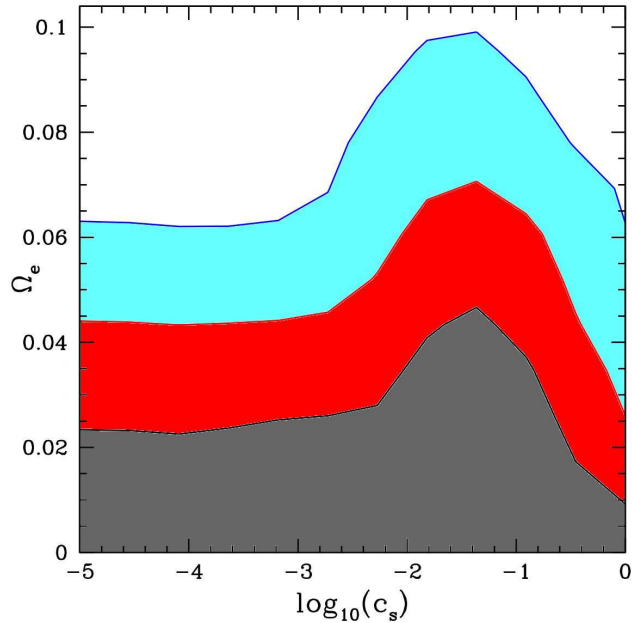,width=3.4in}
  \caption{68.3, 95.4 and 99.7\% confidence level contours in the early 
dark energy model with constant sound speed $c_s$ and early dark energy 
density fraction $\Omega_e$.  The constraints are 
based on current data including CMB, supernovae, LRG power spectrum and 
crosscorrelation of CMB with matter tracers.  From \citet{sound}.} 
  \label{fig:soundw}
  \end{center}
\end{figure}

One quite natural model in this class, and possessing the interesting 
properties of ``predicting'' that $w\approx-1$ today and solving the 
coincidence problem, is the barotropic aether model of \citet{linsch}.  
The constraints on the sound speed, shown in Figure~\ref{fig:soundw}, 
due to current data are not definitive but show a slight preference for 
a low sound speed \citep{sound}.  

Future data using a large galaxy sample for auto- and 
cross-correlations will provide a much clearer picture.  Models with 
extended gravity can also be treated in terms of an effective sound 
speed.  The potential for future data to explore new degrees of freedom 
for dark energy, in terms of sound speed and early dark energy, is quite 
exciting. 

\subsection{Extended Gravity \label{sec:grav}} 

If dark energy is not a new physical component but a modification of 
the equations of motion, e.g.\ gravity beyond general relativity (GR), 
then we need a clear way of parameterizing the changes.  This is 
most commonly accomplished through the relationship between the metric 
potentials $\psi$ and $\phi$ (which are equal within GR) and through 
the form of the Poisson equation or effective gravitational constant. 
See Table~1 of \citet{daniel10} for a translation table among common 
approaches.  

\begin{figure}[!htb]
\begin{center}
\psfig{file=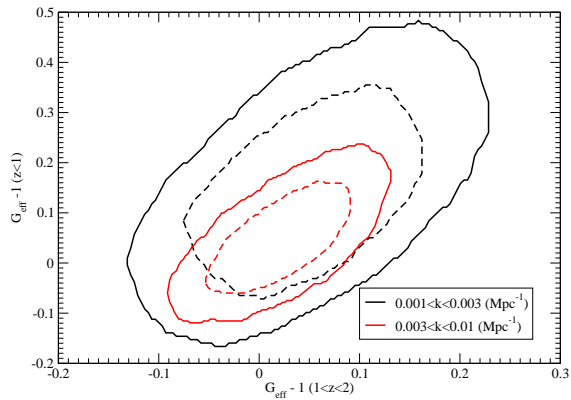,angle=-90,width=3.4in}
\caption{Current data constraints on the fractional deviation 
of the gravitational coupling from Newton's constant, $G_{\rm eff}-1$, 
where this is fit for two bins in redshift $z$ and two bins in 
wavenumber $k$. (Zero deviation is assumed outside the bins.) 
Solid (dashed) contours give 95\% (68\%) cl and are marginalized 
over the other extended gravity parameter $\Sigma$ entering the 
matter perturbation growth equation.  From \citet{daniel10b}. 
}
\label{fig:geffsig}
\end{center}
\end{figure}

Testing gravity on cosmic scales is an area of intense interest at the 
moment; using the most current data \citet{daniel10,bean10,zhao10,uros10} 
find consistency with GR, although again deviations are certainly 
allowed.  See for example Figure~\ref{fig:geffsig}. 

The amplitude of 
the permitted deviations from GR depends on the functional forms assumed 
(e.g.\ time and scale dependence) and the covariances between them.  
Better data from growth probes could play a key role in tightening 
constraints or uncovering new physics.   A particularly exciting 
prospect is comparing the density, velocity, and potential field 
information through combining imaging and spectroscopic surveys 
\citep{jainzhang,uros10,jainkhoury}.

%%%%%%%%%%%%%%%%%%%%%%%%%%%%%%%%%%%%%%%%%%%%%%%%%%%%%%%%%%%%%%
%\section[From Here to Eternity: 50 Ways to Leave $\Lambda$]{From Here 
\section{From Here to Eternity: 50 Ways to Leave $\Lambda$} \label{sec:lam}

To understand the nature of cosmic acceleration, one must know not only 
that $w\approx-1$ today, but the physics behind it.  This is essential 
to comprehend its origin, to understand the areas to probe for distinction 
from $\Lambda$, and to predict the fate of the universe.  We present 
examples of three diverse models that are emphatically not $\Lambda$, 
but give $w\approx-1$, as illustrations of very different physics that 
can cause the behavior to approach that of $\Lambda$.  There are many 
more, and in addition there exist thawing models that for most of the 
history of the universe have $w\approx-1$ but leave $\Lambda$.\footnote{As 
Paul Simon 
almost said: ``The answer is easy if you take it logically / I'd like 
to help you in the puzzle of dark energy / There must be 50 ways to 
leave your Lambda.''} 
These 
cases exemplify that current observations that appear as $w\approx-1$ 
are very far from indicating that dark energy is the cosmological constant. 

The following three model examples are consistent with current data, 
$w\approx-1$ in an averaged sense, yet with completely distinct physics 
from $\Lambda$ and from each other. 
One model looks like $\Lambda$ through microphysics, one through high 
energy physics, and one through gravity.

\subsection{Microphysics \label{sec:micro}} 

The equation of state ratio $w=p/\rho$ is an implicit relation between 
pressure and density but one can instead impose an explicit equation of 
state $p=p(\rho)$.  These are called barotropic fluids and their dynamics 
is determined in terms of their microphysics, e.g.\ the sound speed of 
perturbations $c_s=\sqrt{dp/d\rho}$, by 
\beq 
w'=-3(1+w)(c_s^2-w)\,. \label{eq:barowp}
\eeq 
By inspection this will have an attractor at $w=-1$ and so is an 
attractive class of models for understanding why $w$ may be close to 
$-1$. 

\begin{figure}[!htb]
\begin{center}
\psfig{file=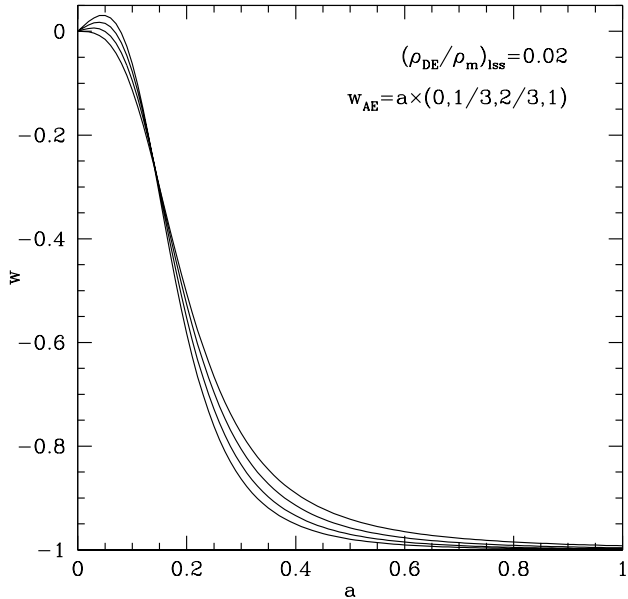,width=3.4in}
\caption{Barotropic models make a rapid transition from $w=0$ at 
high redshift ($a\ll1$) to $w\approx-1$ more recently (the transition 
from $w=-0.1$ to $w=-0.9$ always takes less than 1.5 e-folds).  This is 
forced by the physics of Eq.~(\ref{eq:barowp}) and in distinction 
to quintessence gives a prediction that observations of the recent 
universe should find $w\approx-1$. 
}
\label{fig:baro}
\end{center}
\end{figure}

Using only a stability/causality condition $0\le c_s^2\le1$, and that 
in the past dark energy does not dominate over matter, one finds that 
in the past $w\to0$, $c_s^2\to0$ \citep{linsch}.  In the future, as 
mentioned there is an attractor to $w=-1$.  Thus in both the past and 
the future the model looks like $\Lambda$CDM but can have some deviation 
in between.  Because the dark energy traces the matter in the past, there 
is no fine tuning problem, and because of the rapid evolution between 
asymptotic behaviors, there is no coincidence problem.  Over the more recent 
half of cosmic history, $w\approx-1$ naturally (see Figure~\ref{fig:baro}). 
So this is quite an interesting model.

\subsection{High Energy Physics \label{sec:dbi}} 

Naturalness is an issue with many high energy physics explanations 
of dark energy.  Why is the energy scale associated with the dark 
energy density so small compared to the initial conditions characteristic 
of the high energy early universe?  How is the scale (amplitude) and 
the form of the potential protected against quantum corrections?  
The cosmological constant suffers both problems. 

Some quintessence models avoid the first through attractor solutions 
(e.g.\ inverse power law, exponential, or other tracker potentials 
\citep{ratrap,wett88,zws}), while some avoid the second through 
symmetries (e.g.\ pseudo-Nambu Goldstone boson potentials 
\citep{frie95}).  One model that accomplishes both uses the 
Dirac-Born-Infeld action and its relativistic kinematics.  The DBI 
action 
\beq 
{\mathcal L}=-T\sqrt{1-\dot\phi^2/T}+T-V 
\eeq 
can be thought of as a relativistic generalization of quintessence 
with a Lorentz boost factor $\gamma=(1-\dot\phi^2/T)^{-1/2}$.  This 
action can arise from the world volume traced out by a 3-brane in a 
10-dimensional spacetime.  See \citet{siltong,alisiltong} for more 
details and links to string theory. 

The two functions entering are the brane tension $T(\phi)$ and the 
interaction potential $V(\phi)$, but for our purposes all that will 
be important is that the general dynamics depends primarily on simply 
the asymptotic value of the ratio $V/T$ rather than the specific forms 
of the functions.  As described in \citet{dbi1,dbi2}, several new 
classes of attractors become enabled in the ultrarelativistic limit 
$\gamma\to\infty$, solving the first problem mentioned at the beginning 
of this subsection, while the 
geometric origin of the action protects against the second problem. 

\begin{figure}[!htb]
\begin{center}
\psfig{file=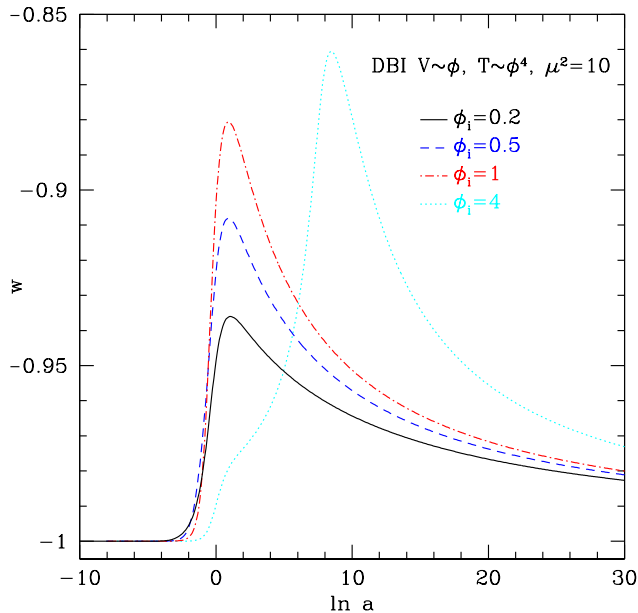,width=3.4in}
\caption{The DBI model has $w\approx-1$ naturally throughout cosmic 
history, for a simple condition on $V/T$.  In the past, the field 
is frozen, then thaws as its dark energy starts to dominate, but 
is attracted back to $w=-1$.  So the deviation from $w=-1$ never 
gets large, despite the radically different physics from a 
cosmological constant vacuum energy.  Adapted from \citet{dbi1}.
}
\label{fig:dbi}
\end{center}
\end{figure}

In particular, an attractor to $w=-1$ exists for $V/T\to\infty$, such 
as for $V\sim\phi^{<2}$ in the simplest $T\sim\phi^4$ case (pure 
AdS$_5$ geometry), or even for the canonical $V=m^2\phi^2$ (mass term), 
$T\sim\phi^4$ case for large mass $m$.  Because the field is frozen 
in the past, and attracted to a $\Lambda$-like state in the future 
(although possessing no nonzero minimum of the potential), then 
$w\approx-1$ always holds.  So this model too agrees with observations 
though originating from very different physics than the cosmological 
constant.  Figure~\ref{fig:dbi} illustrates the concordance.

\subsection{Extended Gravity \label{sec:frgrav}} 

The cosmological constant enters as a constant term added to the 
Ricci scalar in the Einstein-Hilbert action.  Since the spacetime 
curvature plays such an essential role in gravity, it might seem 
strange to add in an independent constant.  Instead, one could 
promote the Ricci scalar to a function, $f(R)$.  For the extensive 
literature on such models, see \citet{durmaar,sotfar,livrel}. 

\begin{figure}[!htb]
\begin{center}
\psfig{file=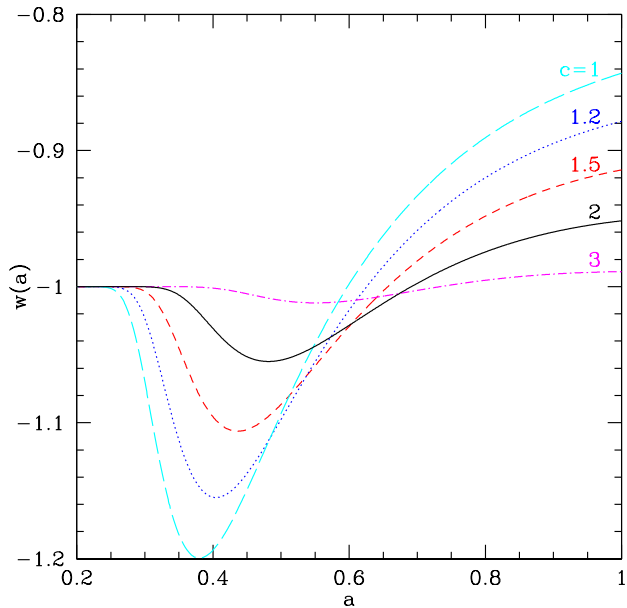,width=3.4in}
\caption{For this extended gravity $f(R)$ model, the effective dark 
energy equation of state naturally has $w\approx-1$ throughout cosmic 
history.  The larger the value of $c$, the more indistinguishable is 
the expansion history from $\Lambda$CDM.  However the growth history 
can have observational signatures.  From \citet{linfr}.
}
\label{fig:frwa}
\end{center}
\end{figure}

Such models not only change the expansion history but add scale 
dependence to spacetime quantities such as light deflection (e.g.\ 
in the parametrized post-Newtonian formalism) as well as the growth of 
density perturbations.  The strong coupling regime in regions of high 
density gradient restore general relativity, but to make this happen 
sufficiently quickly to accord with observations motivates a steep 
dependence on $R$.  Using an additional term in the action of the form 
\beq 
f(R)=-cr\,(1-e^{-R/r})\,, 
\eeq 
where $c$ is a fit parameter and $r$ is determined by $c$ and $\Omega_m$, 
\citet{linfr} found good agreement with both expansion and growth probes. 
(Many other $f(R)$ do also, but this model allows relaxation of fine 
tuning, basically opening up the region $c\lesssim15$.)  
The effective dark energy equation of state (without any physical dark 
energy) possesses $w\approx-1$ for $c>1$, as seen in Figure~\ref{fig:frwa}.

Again, both the past and future appear as a cosmological constant 
universe despite there being no actual cosmological constant. 
The effect of the modified gravity on growth of structure can 
provide observational distinction from $\Lambda$CDM with general 
relativity.  Current measurements, however, are not sufficiently precise 
to impose significant constraints (cf.\ \citet{lucas}, and the model 
independent bounds in Figure~\ref{fig:geffsig}).

%%%%%%%%%%%%%%%%%%%%%%%%%%%%%%%%%%%%%%%%%%%%%%%%%%%%%% 
\section{Conclusions} 

While current data are consistent with a cosmological constant as 
a source for dark energy, a cornucopia of other physical origins 
are in agreement as well.  There are many ways to leave $\Lambda$ 
as an explanation for cosmic acceleration, some without the fine 
tuning and other issues.  We briefly outlined approaches based on 
the microphysical properties of the dark energy, on a high energy 
physics origin, and on extending gravity beyond general relativity. 
All are valid possibilities. 

The exciting goal of future observations is to explore this 
wonderland of physics.  We have seen that for the dynamical aspects, 
next generation measurements of the equation of state and its time 
variation, $w$ and $w'$, in the calibrated form of $w_0$ and $w_a$ 
describe the experimental reach to better than observational accuracy. 
Comparison of tests of growth and expansion could give key clues to 
the underlying physics, as can contrasting the density, velocity, 
and gravitational potential fields of large scale structure.  These 
should be enabled by future wide field imaging and spectroscopic surveys. 

To give a more speculative view, the rich variety of information within 
the CMB, to be revealed by Planck and ground based polarization experiments, 
can explore signatures of early dark energy.  If the early dark energy 
density at CMB last scattering is at much higher levels than the part 
in a billion in the cosmological constant model, then this would be a 
major clue to the physical origin (note percent level contributions can 
be accommodated within the barotropic model of Sec.~\ref{sec:micro}). 
Lensing of the CMB, and weak lensing of galaxies, can probe aspects of 
dark energy clustering and interaction.  Eventually we can hope to have 
as wide an array of aspects of dark energy to probe as have been 
developed for inflation.  We are very much at the beginning of our 
explorations of the physics behind cosmic acceleration.

\section*{Acknowledgments} 

This work has been supported in part by the Director, Office of Science, 
Office of High Energy Physics, of the U.S.\ Department of Energy under 
Contract No.\ DE-AC02-05CH11231, and World Class University grant 
R32-2009-000-10130-0 through the National Research Foundation, Ministry 
of Education, Science and Technology of Korea.

\end{document}